\documentclass[journal]{IEEEtran}

\usepackage{cite}
\usepackage[pdftex]{graphicx}
\usepackage{algorithmic}
\usepackage{array}
\usepackage{stfloats}
\usepackage{booktabs}
\usepackage{url}
\usepackage[]{amsmath}
\usepackage{subfigure}
\usepackage{amssymb,amsfonts,mathrsfs,graphics,color}

\newcommand{\figref}[1]{Fig.\,\ref{#1}}
\newcommand{\tabref}[1]{Table\,\ref{#1}}

\newcommand{\dvec}[2]{
\begin{bmatrix}
      #1\\#2
\end{bmatrix}}
\newcommand{\tvec}[3]{
\begin{bmatrix}
      #1\\#2\\#3
\end{bmatrix}}
\newcommand{\qvec}[4]{
\begin{bmatrix}
      #1\\#2\\#3\\#4
\end{bmatrix}}
\newcommand{\dmat}[4]{
\begin{bmatrix}
      #1 & #2 \\
      #3 & #4
\end{bmatrix}}

\begin{document}

\title{Output Series-Parallel Connection of Passivity-Based Controlled DC-DC Converters: Generalization of Asymptotic Stability}

\author{{Yuma~Murakawa and Takashi~hikihara}
  \thanks{This work was supported in part by the Program on Open Innovation Platform with Enterprises, Research Institute and Academia of Japan Science and Technology Agency, in part by the Cross-ministerial Strategic Innovation Promotion Program No. 18088028 of Japan Science and Technology Agency, and in part by the JSPS KAKENHI under Grant JP20234533.}
  \thanks{The authors are with the Department of Electrical Engineering, Kyoto University, Katsura, Nishikyo-ku, Kyoto, 615-8510 Japan (e-mail: y-murakawa@dove.kuee.kyoto-u.ac.jp; hikihara.takashi.2n@kyoto-u.ac.jp).}
  \thanks{\copyright 2021 IEEE.  Personal use of this material is permitted.  Permission from IEEE must be obtained for all other uses, in any current or future media, including reprinting/republishing this material for advertising or promotional purposes, creating new collective works, for resale or redistribution to servers or lists, or reuse of any copyrighted component of this work in other works.}}

\markboth{~}%
{MURAKAWA \MakeLowercase{\textit{et al.}}:Output Series-Parallel Connection of Passivity-Based Controlled DC-DC Converters: Generalization of Asymptotic Stability}

\maketitle

\begin{abstract}
  The series-paralleling technique of dc-dc converters is utilized in various domains of electrical engineering for improved power conversion. Previous studies have proposed and classified the control schemes for the series-paralleled converters. However, they have several restrictions and lack diversity. The purpose of this paper is to propose passivity-based control (PBC) for the diverse output series-parallel connection of dc-dc converters. It is proved that the output series-paralleled converters regulated by PBC are asymptotically stable. The output series-paralleled converters are numerically simulated to confirm the asymptotic stability. PBC is shown to maintain the stability of the output series-paralleled converters with diverse circuit topologies, parameters, and steady-states. The result of this paper theoretically supports the numerical and experimental considerations in the previous studies and justifies the further extension of the series-parallel connection.
\end{abstract}


\ifCLASSOPTIONpeerreview
  \begin{center} \bfseries EDICS Category: 3-BBND \end{center}
\fi
\IEEEpeerreviewmaketitle

\section{Introduction}

\subsection{Series-Parallel Connection of DC-DC Converters}

\IEEEPARstart{T}{he} series-parallel connection of multiple dc-dc converters is a technique utilized in various fields of electrical engineering \cite{Huang2007, Luo1999, Chen2009, Ruan2020, Walker2004, Pilawa-Podgurski2013, Poshtkouhi2012, Xiong2013, Uzunoglu2009, Blaabjerg2004, Tse2015, Wang2019, Ayad2010, Tofighi2011}. It offers several advantages over a single, high-power, centralized converter such as lower component stresses, increased reliability, ease of maintenance, and improved thermal management \cite{Huang2007, Luo1999, Chen2009, Ruan2020, Walker2004}. In previous researches, this technique has achieved higher efficiency in photovoltaic (PV) systems \cite{Walker2004, Pilawa-Podgurski2013, Poshtkouhi2012}, the load sharing design of high-power converters \cite{Huang2007, Luo1999, Chen2009, Ruan2020}, forming a hybrid power system by combining the distributed power sources \cite{Xiong2013, Uzunoglu2009, Blaabjerg2004, Tse2015, Wang2019, Ayad2010, Tofighi2011}, etc. The series-paralleling technique of dc-dc converters is vital for improved power conversion and the combined use of power sources.

References \cite{Huang2007, Luo1999} have classified the load sharing methods for the paralleled converters. Similarly, series-paralleled converters were investigated in \cite{Chen2009, Ruan2020}. However, these considerations were explicitly aimed to obtain equal load sharing among specified converters. Thus, they cannot be directly applied to the systems which require the arbitrary load sharing among the connection of various converters. These kinds of systems, such as PV systems or hybrid power systems, typically have diverse operating modes depending on the power flow of each converter. In these cases, the control strategies for the converters must be designed to cover the wide range of efficient operation. For example, bifurcations and instabilities of a PV-battery hybrid power system are analyzed in \cite{Xiong2013}. It is necessary to construct a control scheme that widely covers the circuit topologies and the operating range of the dc-dc converters for the diverse series-parallel connection.

\begin{figure}[!t]
  \centering
  \includegraphics[width=0.7\columnwidth]{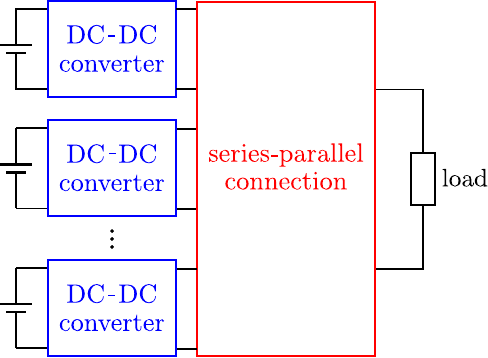}
  \caption{DC-DC converters connected in series-parallel at the output. This paper focuses specifically on the series-parallel connection at the output side.}
  \label{inaat}
\end{figure}

\subsection{Approach Based on Passivity}

There are two categories for the control of dc-dc converters; linear and nonlinear. Previous researches in this area, including the above-mentioned references, have primarily been based on linear control. However, the switched power converters are inherently nonlinear systems \cite{Tse1994, Fossas1996}. Series-paralleled converters also have mutual interactions in contrast to a single dc-dc converter. Hence, linear controllers are unlikely to give robust solutions and optimum performance for the series-paralleled converters \cite{Mazumder2002}. Similarly, eigenvalue and time-delay analyses \cite{Baghaee2017} can reveal the local linearized properties, but due to the nonlinearity of the converters, the performance for the overall operating range of the series-paralleled converters may not be illustrated.

In this study, we adopt passivity-based control (PBC) \cite{Ortega2002, Sira-Ramirez1997, Ortega1998, Sira-Ramirez2006, Ortega2001}, which is a class of a nonlinear control scheme, for the regulation of the series-paralleled converters. It aims to achieve the asymptotic stability of passive systems by adding some damping to the system's storage function.
Since PBC is based on the physical properties of the system, it gives simple and robust control rules \cite{Sira-Ramirez1997, Ortega1998, Ortega2002}.

A system composed of passive subsystems is a passive system \cite{Desoer1969}. A dc-dc converter is also a passive system, in the sense that the switches do not provide any additional energy to the system \cite{Ortega1998, Sira-Ramirez1997}.  They rather serve to intermittently switch the connection structure of the internal passive components \cite{Kassakian1991, Erickson2001}. Hence, the series-paralleled converters constitute a passive system, since each dc-dc converter is a passive subsystem. Then, it is expected from the perspective of energy that the series-paralleled converters are asymptotically stable by applying PBC to each converter. From this viewpoint, several previous research has considered the PBC of series-paralleled converters. The regulation of \`{C}uk converters connected in parallel by PBC is discussed in \cite{Hikihara2011}. Recently, PBC of paralleled boost and buck converters was experimentally verified in \cite{Murakawa2020}. The application of PBC to the hybrid power systems is considered numerically and experimentally in \cite{Ayad2010, Tofighi2011}. Yet, these considerations are limited to specific circuit configurations. Generalized discussions have not been presented for the properties of the series-parallel converters regulated PBC.

\subsection{Contribution and Paper Organization}

In this paper, we theoretically discuss the asymptotic stability of the output series-paralleled dc-dc converters regulated by PBC. The output series-parallel connection of dc-dc converters is shown in \figref{inaat}. This circuit configuration describes systems such as PV systems and hybrid power systems that combine several power sources by the connection of the converters.

Our main contribution of this paper is that we have proved the asymptotic stability of the output series-paralleled converters regulated by PBC. The asymptotic stability is rigorously shown from the perspectives of stored energy and Lyapunov stability theory. We focus on the general converter models and connection structures, in contrast to the studies which consider specific circuit configurations. Hence, the results provided in this paper are general, and are able to support the numerical and experimental results of the previous studies \cite{Hikihara2011, Murakawa2020, Ayad2010, Tofighi2011}. In addition, our theoretical aspects justify the further extension of the series-paralleled converters. In order to verify these features, the numerical simulations of the series-paralleled converters are also provided.

The paper is organized as follows. In section II, we first introduce the general system model of the switched dc-dc converters. Then, the PBC of the dc-dc converters is explained. Some examples are given to show that the model and the PBC can be applied to various types of dc-dc converters. In section III, an external variable representing the mutual interaction is defined to generally describe the output series-parallel connection. The application of PBC to each converter is proved to asymptotically stabilize the series-paralleled converters. The results of the numerical simulations are shown in section IV to confirm the theoretical discussions. Here, several scenarios for external disturbance are considered to verify the stability and the robustness of the output series-paralleled converters regulated by PBC. Finally, section V is the conclusion.

\section{Passivity-Based Control of DC-DC Converters} \label{sec}

In this section, a general model of the dc-dc converters is first introduced. Then, PBC for the dc-dc converters is explained in detail. The control rules for boost, buck, and buck-boost converters are derived as examples. The circuit elements mentioned in this section are assumed ideal.

\begin{table}[t] \caption{List of Symbols} \label{variables}
  \centering
  \renewcommand{\arraystretch}{1.2}
  \scalebox{1.02}{
    \begin{tabular}{lll}
      \hline
      Symbol                                           & Definition         & Property               \\\hline
      $\mathcal{A}\in{\mathbb{R}^{n\times n}}_{>0}$    & Circuit parameter  & Positive definite      \\
      $\mathcal{J}\in\mathbb{R}^{n\times n}$           & Internal structure & Skew-symmetric         \\
      $\mathcal{R}\in{\mathbb{R}^{n\times n}}_{\geq0}$ & Dissipation        & Positive semi-definite \\
      $\boldsymbol{g}\in\mathbb{R}^{n\times m}$        & Input structure    &                        \\
      $\boldsymbol{x}\in\mathbb{R}^n$                  & State variables    &                        \\
      $\boldsymbol{u}\in\mathbb{R}^{m}$                & Input              &                        \\
      $\boldsymbol{s}\in[0,1]^l$                       & Duty ratios        &                        \\
      \hline\end{tabular}}
\end{table}

\begin{figure}[t]
  \centering
  \includegraphics[scale=1.2]{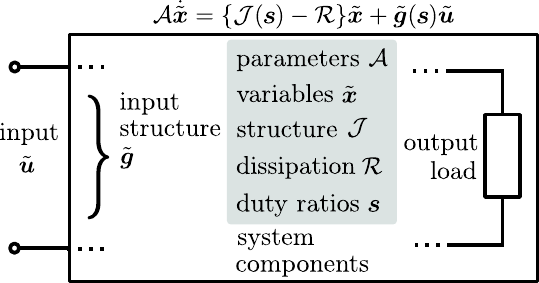}
  \caption{Diagrammatical interpretation of dc-dc converters as port network. Output load is considered as internal element.}
  \label{dcdc_model}
\end{figure}

\subsection{DC-DC Converter Model} \label{model}

As a general model for the dc-dc converters,
\begin{align}
  \mathcal{A}\dot{\boldsymbol{x}}=\{\mathcal{J}(\boldsymbol{s})-\mathcal{R}\}\boldsymbol{x}+\boldsymbol{g}(\boldsymbol{s})\boldsymbol{u} \label{dc}
\end{align}
is assumed. The dot ($\dot{~}$) is a notation for a time differentiation. The symbols in \eqref{dc} are denoted in \tabref{variables}, where $n, m, l\in\mathbb{N}$ are natural numbers. This model is essentially a port-controlled Hamiltonian system \cite{Schaft2000} with minor modifications. Moreover, \eqref{dc} is an averaged model \cite{Sira-Ramirez1997, Ortega1998, Sira-Ramirez2006, Krein1990} based on the assumption that the switches have sufficiently fast switching frequency. Thus, the model is approximated to be a smooth system controlled by the duty ratios $\boldsymbol{s}\in[0,1]^l$. It is confirmed that \eqref{dc} describes a variety of dc-dc converters in \cite{Sira-Ramirez2006, Ortega1998}.

The steady-state equation i.e. the null dynamics of \eqref{dc} is obtained as
\begin{align}
  \{\mathcal{J}(\boldsymbol{s})-\mathcal{R}\}\boldsymbol{x}+\boldsymbol{g}(\boldsymbol{s})\boldsymbol{u}=0 \label{ss}
\end{align}
by setting the differential terms to zero. The desired state $\boldsymbol{x}=\boldsymbol{x}_\mathrm{d}$ and the corresponding duty ratios $\boldsymbol{s}=\boldsymbol{s}_\mathrm{d}$ are determined as the constants to satisfy \eqref{ss}. Hereafter, the subscript `$\mathrm{d}$' indicates the desired value. The control aims to achieve the asymptotic stability at $\boldsymbol{x}=\boldsymbol{x}_\mathrm{d}$ by modifying the duty ratios $\boldsymbol{s}$ at the transient state.

The error $\tilde{\boldsymbol{x}}=\boldsymbol{x}-\boldsymbol{x}_\mathrm{d}$ satisfies
\begin{align}
  \mathcal{A}\dot{\tilde{\boldsymbol{x}}}=\{\mathcal{J}(\boldsymbol{s})-\mathcal{R}\}\tilde{\boldsymbol{x}}+\tilde{\boldsymbol{g}}(\boldsymbol{s})\tilde{\boldsymbol{u}} \label{er}
\end{align}
from \eqref{dc} and \eqref{ss}. The input structure $\tilde{\boldsymbol{g}}$ and input $\tilde{\boldsymbol{u}}$ are defined as
\begin{align}
  \tilde{\boldsymbol{g}}(\boldsymbol{s}) = [\mathcal{J}(\boldsymbol{s})-\mathcal{R}\ {\boldsymbol{g}}(\boldsymbol{s})],\,\tilde{\boldsymbol{u}}=\dvec{\boldsymbol{x}_\mathrm{d}}{\boldsymbol{u}}.
\end{align}
Then, the system is interpreted as a port network shown in \figref{dcdc_model}. The system description of \eqref{er} simplifies the control task to achieving the asymptotic stability at $\tilde{\boldsymbol{x}}=0$. Thus, we consider PBC in the following discussions by \eqref{er}.

\begin{figure}[!t]
  \centering
  \includegraphics[width=1\columnwidth]{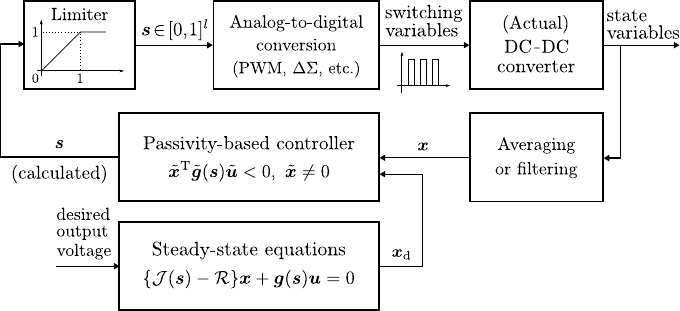}
  \caption{Feedback control scheme for the PBC of dc-dc converters.}
  \label{pbcblockdiagram}
\end{figure}

\subsection{Passivity-Based Control}

The storage function of \eqref{er} is defined as
\begin{align}
  \mathcal{H}=\dfrac{1}{2}\tilde{\boldsymbol{x}}^\mathrm{T}\mathcal{A}\tilde{\boldsymbol{x}}. \label{h}
\end{align}
\eqref{h} has its minimum at $\tilde{\boldsymbol{x}}=0$. The time derivative of the storage function \eqref{h} is
\begin{align}
  \dot{\mathcal{H}} & =\tilde{\boldsymbol{x}}^\mathrm{T}\mathcal{A}\dot{\tilde{\boldsymbol{x}}} \nonumber                                                                                               \\
                    & =-\tilde{\boldsymbol{x}}^\mathrm{T}\mathcal{R}\tilde{\boldsymbol{x}} +\tilde{\boldsymbol{x}}^\mathrm{T}\tilde{\boldsymbol{g}}(\boldsymbol{s})\tilde{\boldsymbol{u}}. \label{dhdt}
\end{align}
Here, we assume $\dot{\boldsymbol{x}}_\mathrm{d}=0$, since we are considering a dc system. This assumption yields simpler discussions and control compared to \cite{Sira-Ramirez1997, Ortega1998, Sira-Ramirez2006}. The first term on the right-hand side $-\tilde{\boldsymbol{x}}^\mathrm{T}\mathcal{R}\tilde{\boldsymbol{x}}$ is the dissipation of the system, which is negative at $\tilde{\boldsymbol{x}}\neq0$. The second term $\tilde{\boldsymbol{x}}^\mathrm{T}\tilde{\boldsymbol{g}}(\boldsymbol{s})\tilde{\boldsymbol{u}}$ corresponds to the power supplied from the externals.

Therefore, when the externally supplied power satisfies
\begin{align}
  \tilde{\boldsymbol{x}}^\mathrm{T}\tilde{\boldsymbol{g}}(\boldsymbol{s})\tilde{\boldsymbol{u}}<0,\ \tilde{\boldsymbol{x}}\neq0, \label{cond}
\end{align}
the remaining energy dissipation guarantees
\begin{align}
  \dot{\mathcal{H}}=-\tilde{\boldsymbol{x}}^\mathrm{T}\mathcal{R}\tilde{\boldsymbol{x}} +\tilde{\boldsymbol{x}}^\mathrm{T}\tilde{\boldsymbol{g}}(\boldsymbol{s})\tilde{\boldsymbol{u}}<0,\
  \tilde{\boldsymbol{x}}\neq0. \label{lya}
\end{align}
The storage function $\mathcal{H}$ plays the role of a Lyapunov function of \eqref{er} and ensures the asymptotic stability at $\tilde{\boldsymbol{x}}=0$. To summarize, PBC for the dc-dc converters prepares a control rule for the duty ratios $\boldsymbol{s}$ given to satisfy the condition \eqref{cond}.

\begin{figure}[!t]
  \centering
  \subfigure[~]{
    \includegraphics[scale=1]{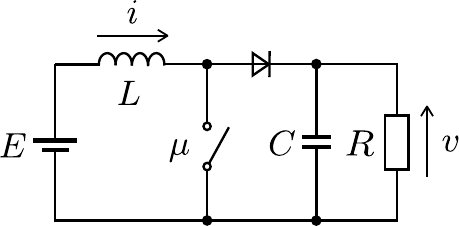}\label{boostfig}}\\[3pt]
  \subfigure[~]{
    \includegraphics[scale=1]{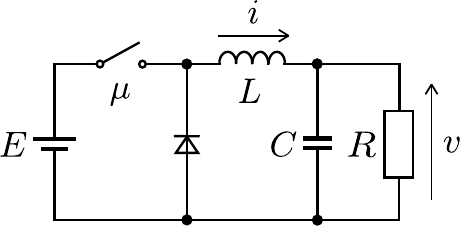}\label{buckfig}}\\[9pt]
  \subfigure[~]{
    \includegraphics[scale=1]{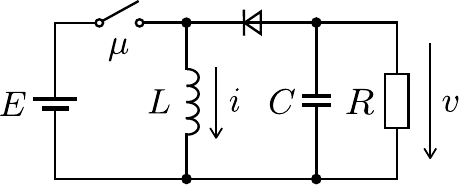}\label{buckboostfig}}
  \caption{Example circuit schematics of dc-dc converters. (a) Boost converter. (b) Buck converter. (c) Buck-boost converter.}
\end{figure}

\figref{pbcblockdiagram} shows the schematic diagram of the feedback control scheme for the PBC of dc-dc converters. The actual dc-dc converter is regulated by a discrete switching variable, hence the duty ratio is processed through analog-to-digital conversion, typically pulse-width modulation (PWM) or delta-sigma modulation \cite{Sira-Ramirez2006, Silva-Ortigoza2014}. A hard limiter is implemented to guarantee the duty ratio to be in the closed interval [0,1]. Regardless, the feedback gain should be designed for the duty ratios not to exceed the interval while the converter is operating in the expected region. Averaging or filtering is also necessary to adjust the state variables for the feedback control based on the averaged system. The actual state variables remain small high-frequency ripples, due to the switching operation. However, this process is usually neglected, owing to the low-pass filtering nature of the passive elements and the sensors. Typically, the desired state ${\boldsymbol{x}}_\mathrm{d}$ is uniquely determined by \eqref{ss} to obtain a given desired output voltage.

\subsection{Examples of PBC for DC-DC Converters}

\subsubsection{Boost Converter} \figref{boostfig} is the circuit schematic of the boost converter. The boost converter is modeled as
\begin{align}
  \dmat{L}{0}{0}{C}\dvec{\dot{i}}{\dot{v}}=\dmat{0}{-(1-\mu)}{(1-\mu)}{-1/R}\dvec{i}{v}+\dvec{1}{0}E. \label{boostde}
\end{align}
The steady-state equations are obtained as
\begin{align}
  \begin{cases}
    \mu=1 - \dfrac{E}{v}, \\[3pt]
    Ei=\dfrac{v^2}{R}.
  \end{cases}\label{boostss}
\end{align}
The desired state $[i\,v]^\mathrm{T}=[i_\mathrm{d}\,v_\mathrm{d}]^\mathrm{T}$ and the desired duty ratio $\mu_\mathrm{d}\in[0,1]$ are determined as constants to satisfy \eqref{boostss}.

Putting $\tilde{\boldsymbol{x}}=[i-i_\mathrm{d}\,v-v_\mathrm{d}]^\mathrm{T}$ gives the storage function of
\begin{align}
  \mathcal{H}= & \dfrac{1}{2}\tilde{\boldsymbol{x}}^\mathrm{T}\mathcal{A}\tilde{\boldsymbol{x}}=\dfrac{1}{2}\dvec{i-i_\mathrm{d}}{v-v_\mathrm{d}}^\mathrm{T}\dmat{L}{0}{0}{C}\dvec{i-i_\mathrm{d}}{v-v_\mathrm{d}} \nonumber \\[1pt]
  =            & \dfrac{1}{2}L(i-i_\mathrm{d})^2+\dfrac{1}{2}C(v-v_\mathrm{d})^2. \label{boosth}
\end{align}
In order to form \eqref{boosth} as a Lyapunov function,
\begin{align}
  \tilde{\boldsymbol{x}}^\mathrm{T}\tilde{\boldsymbol{g}}(\mu)\tilde{\boldsymbol{u}} & =
  \dvec{i-i_\mathrm{d}}{v-v_\mathrm{d}}^\mathrm{T}
  \begin{bmatrix}
    0 & -(1-\mu) & 1 \\ (1-\mu)& -1/R & 0
  \end{bmatrix}
  \tvec{i_\mathrm{d}}{v_\mathrm{d}}{E}\nonumber                                                                                                                                              \\
                                                                                     & =(\mu- \mu_\mathrm{d})(iv_\mathrm{d}-i_\mathrm{d}v)<0,\ \tilde{\boldsymbol{x}}\neq0 \label{boostcond}
\end{align}
has to be satisfied according to \eqref{cond}. The condition \eqref{boostcond} is fulfilled by controlling the duty ratio $\mu$ as
\begin{align}
  \mu = \mu_\mathrm{d}-k(iv_\mathrm{d}-i_\mathrm{d}v),\ k>0, \label{boostpbc}
\end{align}
where $k$ is a positive feedback gain. Therefore, PBC for the boost converter is obtained as \eqref{boostpbc}.

\subsubsection{Buck Converter} PBC for the buck converter follows the same derivation process as the boost converter. The buck converter shown in \figref{buckfig} is modeled as
\begin{align}
  \dmat{L}{0}{0}{C}\dvec{\dot{i}}{\dot{v}}=\dmat{0}{-1}{1}{-1/R}\dvec{i}{v}+\dvec{\mu}{0}E. \label{buckde}
\end{align}
The condition for the asymptotic stability is
\begin{align}
  \tilde{\boldsymbol{x}}^\mathrm{T}\tilde{\boldsymbol{g}}(\mu)\tilde{\boldsymbol{u}} & =
  \dvec{i-i_\mathrm{d}}{v-v_\mathrm{d}}^\mathrm{T}
  \begin{bmatrix}
    0 & -1 & \mu \\ 1& -1/R & 0
  \end{bmatrix}
  \tvec{i_\mathrm{d}}{v_\mathrm{d}}{E}\nonumber                                                                                                                 \\
                                                                                     & =(\mu- \mu_\mathrm{d})(i-i_\mathrm{d})E<0,\ \tilde{\boldsymbol{x}}\neq0,
\end{align}
from \eqref{cond}. This condition is satisfied by
\begin{align}
  \mu = \mu_\mathrm{d}-k(i-i_\mathrm{d}),\ k>0. \label{buckpbc}
\end{align}
Thus, \eqref{buckpbc} is the PBC for the buck converter.

\subsubsection{Buck-Boost Converter}

PBC for the buck-boost converter also follows the same derivation process. The buck-boost converter shown in \figref{buckboostfig} is modeled as
\begin{align}
  \dmat{L}{0}{0}{C}\dvec{\dot{i}}{\dot{v}}=\dmat{0}{-(1-\mu)}{(1-\mu)}{-1/R}\dvec{i}{v}+\dvec{\mu}{0}E, \label{buckboostde}
\end{align}
The condition for the asymptotic stability is
\begin{align}
  \tilde{\boldsymbol{x}}^\mathrm{T}\tilde{\boldsymbol{g}}(\mu)\tilde{\boldsymbol{u}}=
  \dvec{i-i_\mathrm{d}}{v-v_\mathrm{d}}^\mathrm{T}
  \begin{bmatrix}
    0 & -(1-\mu) & \mu \\ (1-\mu)& -1/R & 0
  \end{bmatrix}
  \tvec{i_\mathrm{d}}{v_\mathrm{d}}{E}\nonumber \\[2pt]
  =(\mu- \mu_\mathrm{d})\{i(v_\mathrm{d}+E)-i_\mathrm{d}(v+E)\}<0,\ \tilde{\boldsymbol{x}}\neq0. \label{buckboostcond}
\end{align}
This condition is satisfied by
\begin{align}
  \mu = \mu_\mathrm{d}-k\{i(v_\mathrm{d}+E)-i_\mathrm{d}(v+E)\},\ k>0. \label{buckboostpbc}
\end{align}
Thus, \eqref{buckboostpbc} is the PBC for the buck-boost converter.

\section[Output Series-Parallel Connection of DC-DC Converters]{Output Series-Parallel \\ Connection of DC-DC Converters}

In this section, we prove that the output series-paralleled converters regulated by PBC are asymptotically stable. First, the dc-dc converter model is introduced with an additional external variable representing the mutual interaction between the converters. Then, certain circuit restrictions are given to the variables in order for the model to describe a series or parallel connection. Through the process, the output series-paralleled converters are shown to be reclassified into the general dc-dc converter model of \eqref{er}. Thus, the asymptotic stability of the series-paralleled converters can be discussed based on \eqref{cond}. Finally, it is proved that the dc-dc converters regulated by PBC can be connected in series-parallel at the output while maintaining their asymptotic stability.

\begin{figure}[!t]
  \centering
  \hfill\includegraphics[width=0.95\columnwidth]{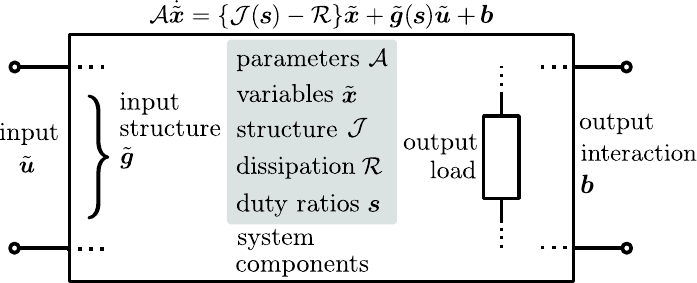}
  \caption{DC-DC converter model with interaction at output.}
  \label{interconnected_model}
  \vspace{15pt}
  \subfigure[~]{
    \includegraphics[scale=1.12]{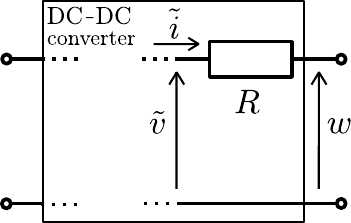}\label{seriesint}}\hfill
  \subfigure[~]{
    \includegraphics[scale=1.12]{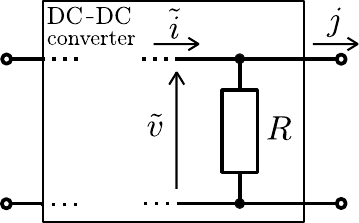}\label{parallelint}}
  \caption{Possible circuit structures for series or parallel output interaction. (a) Series interaction. (b) Parallel interaction.} \label{intent}
\end{figure}

\subsection{DC-DC Converter with Output Interaction}

The dc-dc converter with output interaction is modeled as
\begin{align}
  \mathcal{A}\dot{\tilde{\boldsymbol{x}}}=\{\mathcal{J}(\boldsymbol{s})-\mathcal{R}\}\tilde{\boldsymbol{x}}+\tilde{\boldsymbol{g}}(\boldsymbol{s})\tilde{\boldsymbol{u}}+\boldsymbol{b}, \label{ic}
\end{align}
where $\boldsymbol{b}\in\mathbb{R}^n$ implies the interaction at the output. This dc-dc converter model is shown diagrammatically in \figref{interconnected_model}. As shown in the figure, the interaction $\boldsymbol{b}$ is treated as an external variable. The storage function of \eqref{ic} is kept as
\begin{align}
  \mathcal{H}=\dfrac{1}{2}\tilde{\boldsymbol{x}}^\mathrm{T}\mathcal{A}\tilde{\boldsymbol{x}}. \label{h2}
\end{align}
However, its time derivative is altered to be
\begin{align}
  \dot{\mathcal{H}} & =\tilde{\boldsymbol{x}}^\mathrm{T}\mathcal{A}\dot{\tilde{\boldsymbol{x}}} \nonumber                                                                                                                                                \\
                    & =-\tilde{\boldsymbol{x}}^\mathrm{T}\mathcal{R}\tilde{\boldsymbol{x}} +\tilde{\boldsymbol{x}}^\mathrm{T}\tilde{\boldsymbol{g}}(\boldsymbol{s})\tilde{\boldsymbol{u}}+\tilde{\boldsymbol{x}}^\mathrm{T}\boldsymbol{b}, \label{dhdt2}
\end{align}
where the term $\tilde{\boldsymbol{x}}^\mathrm{T}\boldsymbol{b}$ implies the power supplied from the output interaction.

Two possible circuit structures are assumed for the output interaction; series or parallel connection as shown in Figs.\,\ref{seriesint} and (b), respectively. The variables $\tilde{v}$ and $\tilde{i}$ represent the output voltage and the output current of the converter. In \figref{seriesint}, the interaction appears as the voltage $w$ in series to the load. From the assumption that the only dissipative element of the converter is the output load $R$, we have
\begin{align}
  -\tilde{\boldsymbol{x}}^\mathrm{T}\mathcal{R}\tilde{\boldsymbol{x}}=-\dfrac{\tilde{v}^2}{R},\
  \tilde{\boldsymbol{x}}^\mathrm{T}\boldsymbol{b} = \dfrac{\tilde{v}w}{R}\label{seriesinten}
\end{align}
for the series interaction. Clearly, adding the two equations in \eqref{seriesinten} gives the power consumption at the load. Here, we can see that the output interaction affects the dissipation of the model. Similarly, we have
\begin{align}
  -\tilde{\boldsymbol{x}}^\mathrm{T}\mathcal{R}\tilde{\boldsymbol{x}}=-{R}\tilde{i}^2,\
  \tilde{\boldsymbol{x}}^\mathrm{T}\boldsymbol{b} = {R}\tilde{i}j\label{parallelinten}
\end{align}
for the parallel case shown in \figref{parallelint}.

\begin{figure}[!t]
  \centering
  \includegraphics[width=1\columnwidth]{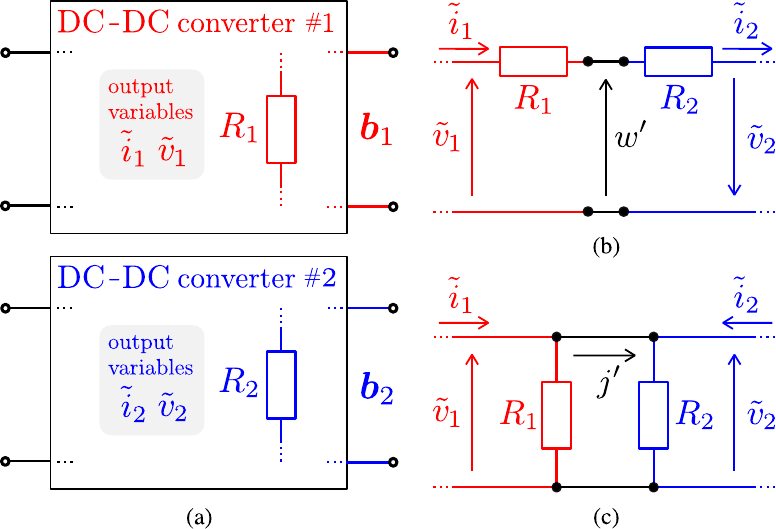}
  \caption{Pair of dc-dc converters with output interaction. (a) Drawn individually. (b) Series connection at output. (c) Parallel connection at output.}
  \label{int}
\end{figure}

\subsection{Output Series-Parallel Connection for Pair of Converters} \label{pair2}

\subsubsection{System Model and Energy Property}

\figref{int}(a) shows a pair of dc-dc converters with output interactions. Each converter has  the output load $R$, the output voltage $\tilde{v}$, the output current $\tilde{i}$, and the output interaction $\boldsymbol{b}$. The subscripts `1' and `2' correspond to converters \#1 and \#2, respectively. These subscripts do not alter the definitions or the properties of the symbols listed in \tabref{variables}. They are used only to distinguish the matrices.

The pair of dc-dc converters shown in \figref{int}(a) are modeled as
\begin{align}
  \begin{cases}
    \mathcal{A}_{1}\dot{\tilde{\boldsymbol{x}}}_{1}=\{\mathcal{J}_{1}(\boldsymbol{s}_{1})-\mathcal{R}_{1}\}\tilde{\boldsymbol{x}}_{1}+\tilde{\boldsymbol{g}}_{1}(\boldsymbol{s}_{1})\tilde{\boldsymbol{u}}_{1}+\boldsymbol{b}_{1}, \\
    \mathcal{A}_{2}\dot{\tilde{\boldsymbol{x}}}_{2}=\{\mathcal{J}_{2}(\boldsymbol{s}_{2})-\mathcal{R}_{2}\}\tilde{\boldsymbol{x}}_{2}+\tilde{\boldsymbol{g}}_{2}(\boldsymbol{s}_{2})\tilde{\boldsymbol{u}}_{2}+\boldsymbol{b}_{2}.
  \end{cases} \label{ic222}
\end{align}
Note that these equations are independent due to the fact that no conditions are fixed to the variables yet. Equation \eqref{ic222} is rewritten as
\begin{align}
  \!\!\mathcal{A}_{12}\dot{\tilde{\boldsymbol{x}}}_{12}=\{\mathcal{J}_{12}(\boldsymbol{s}_{12})
  -\mathcal{R}'_{12}\}\tilde{\boldsymbol{x}}_{12}+\tilde{\boldsymbol{g}}_{12}(\boldsymbol{s}_{12})\tilde{\boldsymbol{u}}_{12}+\boldsymbol{b}_{12},\label{ic2}
\end{align}
by using
\begin{align}\begin{cases}
    \tilde{\boldsymbol{x}}_{12}=\dvec{\tilde{\boldsymbol{x}}_{1}}{\tilde{\boldsymbol{x}}_{2}}\!,\,
    \tilde{\boldsymbol{u}}_{12}=\dvec{\tilde{\boldsymbol{u}}_{1}}{\tilde{\boldsymbol{u}}_{2}}\!,\,
    \mathcal{A}_{12}=\dmat{\mathcal{A}_{1}}{0}{0}{\mathcal{A}_{2}}\!,\,
    \\[13pt]
    \mathcal{J}_{12}=\dmat{\mathcal{J}_{1}}{0}{0}{\mathcal{J}_{2}}\!,\,
    \mathcal{R}'_{12}=\dmat{\mathcal{R}_{1}}{0}{0}{\mathcal{R}_{2}}\!,\,
    \\[13pt]
    \tilde{\boldsymbol{g}}_{12}=\dmat{\tilde{\boldsymbol{g}}_{1}}{0}{0}{\tilde{\boldsymbol{g}}_{2}}\!,\,
    {\boldsymbol{b}}_{12}=\dvec{{\boldsymbol{b}}_{1}}{{\boldsymbol{b}}_{2}}\!,\,
    {\boldsymbol{s}}_{12}=\dvec{{\boldsymbol{s}}_{1}}{{\boldsymbol{s}}_{2}}\!.\,
  \end{cases}\label{param}
\end{align}
Then, the storage function of \eqref{ic2} is
\begin{align}
  \mathcal{H}_{12}=\dfrac{1}{2}{\tilde{\boldsymbol{x}}_{12}}^\mathrm{T}\mathcal{A}_{12}\tilde{\boldsymbol{x}}_{12}.
  \label{ic2h}
\end{align}
The time derivative of the storage function is obtained as
\begin{align}
  \dot{\mathcal{H}}_{12} & = {\tilde{\boldsymbol{x}}_{12}}^\mathrm{T}\mathcal{A}_{12}\dot{\tilde{\boldsymbol{x}}}_{12} \nonumber \\
                         & =-{\tilde{\boldsymbol{x}}_{12}}^\mathrm{T}\mathcal{R}'_{12}\tilde{\boldsymbol{x}}_{12} +
  {\tilde{\boldsymbol{x}}_{12}}^\mathrm{T}\tilde{\boldsymbol{g}}_{12}({\boldsymbol{s}}_{12})\tilde{\boldsymbol{u}}_{12}+{\tilde{\boldsymbol{x}}_{12}}^\mathrm{T}\boldsymbol{b}_{12},
  \label{ic2dhdt}
\end{align}
where
\begin{align}
  \begin{cases}
    {\tilde{\boldsymbol{x}}_{12}}^\mathrm{T}\mathcal{R}'_{12}\tilde{\boldsymbol{x}}_{12}=
      {\tilde{\boldsymbol{x}}_{1}}^\mathrm{T}\mathcal{R}_{1}\tilde{\boldsymbol{x}}_{1}+
    {\tilde{\boldsymbol{x}}_{2}}^\mathrm{T}\mathcal{R}_{2}\tilde{\boldsymbol{x}}_{2}, \\[2pt]
    {\tilde{\boldsymbol{x}}_{12}}^\mathrm{T}\boldsymbol{b}_{12} = {\tilde{\boldsymbol{x}}_{1}}^\mathrm{T}\boldsymbol{b}_{1}+{\tilde{\boldsymbol{x}}_{2}}^\mathrm{T}\boldsymbol{b}_{2}
  \end{cases}
  \label{28}
\end{align}
from \eqref{param}.

\subsubsection{Series Connection} \label{sericon}

\figref{int}(b) is the schematic of the outputs connected in series. In this configuration, both loads have the identical current flow. The converters interact through the voltage $w'$, which represents the voltage imbalance among the outputs. Therefore, we have
\begin{align}\tilde{i}_1=\tilde{i}_2,\ w_1=-w_2,\label{current}\end{align}
where $w_1$ and $w_2$ indicates the voltage interaction for converters \#1 and \#2, respectively.
Also from the voltage law,
\begin{align}w_1=\tilde{v}_1-R_1\tilde{i}_1,\ w_2=\tilde{v}_2-R_2\tilde{i}_2\label{kvl}\end{align}
is obtained.
Then, the load current is
\begin{align}\tilde{i}_1=\tilde{i}_2=\dfrac{\tilde{v}_1+\tilde{v}_2}{R_1+R_2}\label{ccurrent}\end{align}
from \eqref{current} and \eqref{kvl}.

Since \figref{int}(b) corresponds to the series interaction shown in \figref{seriesint}, we have
\begin{align}
  \begin{cases}
    {\tilde{\boldsymbol{x}}_{1}}^\mathrm{T}\mathcal{R}_{1}\tilde{\boldsymbol{x}}_{1}=\dfrac{{\tilde{v}_1}{}^2}{R_1},\
    {\tilde{\boldsymbol{x}}_{2}}^\mathrm{T}\mathcal{R}_{2}\tilde{\boldsymbol{x}}_{2}=\dfrac{{\tilde{v}_2}{}^2}{R_2}, \\[3pt]
    {\tilde{\boldsymbol{x}}_{1}}^\mathrm{T}\boldsymbol{b}_{1}=\dfrac{\tilde{v}_1w_1}{R_1},\
    {\tilde{\boldsymbol{x}}_{2}}^\mathrm{T}\boldsymbol{b}_{2}=\dfrac{\tilde{v}_2w_2}{R_2}
  \end{cases}
  \label{serienergy}
\end{align}
from \eqref{seriesinten}. Moreover, from \eqref{kvl} and \eqref{ccurrent}, we have
\begin{align}
  \begin{cases}
    {\tilde{\boldsymbol{x}}_{1}}^\mathrm{T}\boldsymbol{b}_{1}=\dfrac{{\tilde{v}_1}{}^2}{R_1}-\dfrac{\tilde{v}_1(\tilde{v}_1+\tilde{v}_2)}{R_1+R_2}, \\[8pt]
    {\tilde{\boldsymbol{x}}_{2}}^\mathrm{T}\boldsymbol{b}_{2}=\dfrac{{\tilde{v}_2}{}^2}{R_2}-\dfrac{\tilde{v}_2(\tilde{v}_1+\tilde{v}_2)}{R_1+R_2}.
  \end{cases}\label{seri0}
\end{align}
Substituting \eqref{seri0} to \eqref{28} results in
\begin{align}
  {\tilde{\boldsymbol{x}}_{12}}^\mathrm{T}\boldsymbol{b}_{12}
   & =\dfrac{{\tilde{v}_1}{}^2}{R_1}+\dfrac{{\tilde{v}_2}{}^2}{R_2}-\dfrac{(\tilde{v}_1+\tilde{v}_2)^2}{R_1+R_2}\nonumber \\[1pt]
   & ={\tilde{\boldsymbol{x}}_{12}}^\mathrm{T}\mathcal{R}'_{12}\tilde{\boldsymbol{x}}_{12}
  -{\tilde{\boldsymbol{x}}_{12}}^\mathrm{T}\mathcal{R}_{\mathrm{s}}\tilde{\boldsymbol{x}}_{12},
\end{align}
by introducing a positive semi-definite matrix $\mathcal{R}_{\mathrm{s}}$ that satisfies
\begin{align}
  {\tilde{\boldsymbol{x}}_{12}}^\mathrm{T}\mathcal{R}_{\mathrm{s}}\tilde{\boldsymbol{x}}_{12}=\dfrac{(\tilde{v}_1+\tilde{v}_2)^2}{R_1+R_2}.
\end{align}
Hence, we obtain
\begin{align}
  \boldsymbol{b}_{12}
  =\mathcal{R}'_{12}\tilde{\boldsymbol{x}}_{12}
  -\mathcal{R}_{\mathrm{s}}\tilde{\boldsymbol{x}}_{12} \label{serib}
\end{align}
for the series connection in \figref{int}(b). From above discussions, we can confirm that the connection at the output alters the dissipation of the model.

\subsubsection{Parallel Connection} \label{paracon}

Similar discussions can be made for the parallel connection. \figref{int}(c) is the schematic of the outputs connected in parallel. In this configuration, both loads have identical load voltage. The current $j'$ implies the interaction between the converters. Therefore, we have
\begin{align}\tilde{v}_1=\tilde{v}_2,\ j_1=-j_2,\label{voltage}\end{align}
where $j_1$ and $j_2$ indicates the current interaction for converters \#1 and \#2, respectively. Also from the current law,
\begin{align}j_1=\tilde{i}_1-\dfrac{\tilde{v}_1}{R_1},\ j_2=\tilde{i}_2-\dfrac{\tilde{v}_2}{R_2}\label{kcl}\end{align}
is obtained.
Then, the load voltage is
\begin{align}\tilde{v}_1=\tilde{v}_2=\dfrac{R_1R_2}{R_1+R_2}(\tilde{i}_1+\tilde{i}_2)\label{vvoltage}\end{align}
from \eqref{voltage} and \eqref{kcl}.

Since \figref{int}(c) corresponds to the series interaction shown in \figref{parallelint}, we have
\begin{align}
  \begin{cases}
    {\tilde{\boldsymbol{x}}_{1}}^\mathrm{T}\mathcal{R}_{1}\tilde{\boldsymbol{x}}_{1}={R_1}{\tilde{i}_1}{}^2,\
    {\tilde{\boldsymbol{x}}_{2}}^\mathrm{T}\mathcal{R}_{2}\tilde{\boldsymbol{x}}_{2}={R_2}{\tilde{i}_2}{}^2, \\[2pt]
    {\tilde{\boldsymbol{x}}_{1}}^\mathrm{T}\boldsymbol{b}_{1}={R_1}{\tilde{i}_1}j_1,\
    {\tilde{\boldsymbol{x}}_{2}}^\mathrm{T}\boldsymbol{b}_{2}={R_2}{\tilde{i}_2}j_2
  \end{cases}
  \label{paraenergy}
\end{align}
from \eqref{parallelinten}. Moreover, from \eqref{kcl} and \eqref{vvoltage}, we have
\begin{align}
  \begin{cases}
    {\tilde{\boldsymbol{x}}_{1}}^\mathrm{T}\boldsymbol{b}_{1}={R_1}{\tilde{i}_1}{}^2-\dfrac{R_1R_2}{R_1+R_2}\tilde{i}_1(\tilde{i}_1+\tilde{i}_2), \\[7pt]
    {\tilde{\boldsymbol{x}}_{2}}^\mathrm{T}\boldsymbol{b}_{2}={R_2}{\tilde{i}_2}{}^2-\dfrac{R_1R_2}{R_1+R_2}\tilde{i}_2(\tilde{i}_1+\tilde{i}_2).
  \end{cases}\label{para0}
\end{align}
Substituting \eqref{para0} to \eqref{28} results in
\begin{align}
  {\tilde{\boldsymbol{x}}_{12}}^\mathrm{T}\boldsymbol{b}_{12}
   & ={R_1}{\tilde{i}_1}{}^2+{R_2}{\tilde{i}_2}{}^2-\dfrac{R_1R_2}{R_1+R_2}(\tilde{i}_1+\tilde{i}_2)^2\nonumber \\[2pt]
   & ={\tilde{\boldsymbol{x}}_{12}}^\mathrm{T}\mathcal{R}'_{12}\tilde{\boldsymbol{x}}_{12}
  -{\tilde{\boldsymbol{x}}_{12}}^\mathrm{T}\mathcal{R}_{\mathrm{p}}\tilde{\boldsymbol{x}}_{12},
\end{align}
by introducing a positive semi-definite matrix $\mathcal{R}_{\mathrm{p}}$ that satisfies
\begin{align}
  {\tilde{\boldsymbol{x}}_{12}}^\mathrm{T}\mathcal{R}_{\mathrm{p}}\tilde{\boldsymbol{x}}_{12}=\dfrac{R_1R_2}{R_1+R_2}(\tilde{i}_1+\tilde{i}_2)^2.
\end{align}
Hence, we obtain
\begin{align}
  \boldsymbol{b}_{12}
  =\mathcal{R}'_{12}\tilde{\boldsymbol{x}}_{12}
  -\mathcal{R}_{\mathrm{p}}\tilde{\boldsymbol{x}}_{12} \label{parab}
\end{align}
for the parallel connection in \figref{int}(c).

\subsubsection{Replacement of Dissipation} \label{sym}

From \eqref{serib} and \eqref{parab}, both series and parallel interactions at the output replace the original dissipation matrix. It is possible to rewrite the model of \eqref{ic2} as
\begin{align}
  \mathcal{A}_{12}\dot{\tilde{\boldsymbol{x}}}_{12}=\{\mathcal{J}_{12}(\boldsymbol{s}_{12})
  -\mathcal{R}_{12}\}\tilde{\boldsymbol{x}}_{12}+\tilde{\boldsymbol{g}}_{12}(\boldsymbol{s}_{12})\tilde{\boldsymbol{u}}_{12}\label{ic3}
\end{align}
by substituting \eqref{serib} or \eqref{parab} to \eqref{ic2}. Here, $\mathcal{R}_{12}$ is equal to $\mathcal{R}_{\mathrm{s}}$ or $\mathcal{R}_{\mathrm{p}}$ depending on the series or parallel connection at the output. The model \eqref{ic3} has the same structure as \eqref{er}, which is the general model of the dc-dc converter. Therefore, we have shown that a pair of dc-dc converters connected in either series or parallel at the output is described by the general converter model of \eqref{er}.

\begin{figure}[!t]
  \centering
  \includegraphics[width=1\columnwidth]{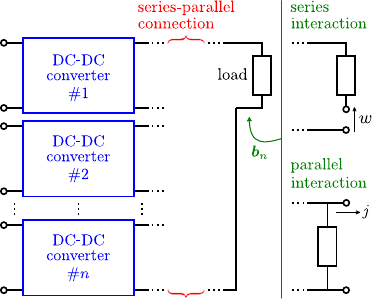}
  \caption{Output series-paralleled $n$ converters.}
  \label{SPn}
\end{figure}

\subsection[Output Series-Parallel Connection of n Converters]{Output Series-Parallel Connection of $n$ Converters} \label{outputsp}

Let us first clarify that the system consisting of $n\in \mathbb{N}$ numbers of converters connected repeatedly in either series or parallel at their outputs to share a single load is defined as output series-paralleled $n$ converters. \figref{SPn} shows the diagram of the output series-paralleled $n$ converters. The output interaction for this system is also given in series or parallel to the output load, similarly to \figref{intent}. The single dc-dc converter is included in the definition for the case of $n=1$.

Suppose, for all $m\in\mathbb{N}\ \mathrm{s.t.}\ m\leq n$, any output series-paralleled $m$ converters are described by \eqref{er}. Consider the natural numbers $\alpha,\,\beta\in\mathbb{N}\ \mathrm{s.t.}\ \alpha+\beta=n+1$. Then, a specific output series-paralleled $n+1$ converters can be described by the model
\begin{align}
  \begin{cases}
    \mathcal{A}_{\alpha}\dot{\tilde{\boldsymbol{x}}}_{\alpha}=\{\mathcal{J}_{\alpha}(\boldsymbol{s}_{\alpha})-\mathcal{R}_{\alpha}\}\tilde{\boldsymbol{x}}_{\alpha}+\tilde{\boldsymbol{g}}_{\alpha}(\boldsymbol{s}_{\alpha})\tilde{\boldsymbol{u}}_{\alpha}+\boldsymbol{b}_{\alpha}, \\
    \mathcal{A}_{\beta}\dot{\tilde{\boldsymbol{x}}}_{\beta}=\{\mathcal{J}_{\beta}(\boldsymbol{s}_{\beta})-\mathcal{R}_{\beta}\}\tilde{\boldsymbol{x}}_{\beta}+\tilde{\boldsymbol{g}}_{\beta}(\boldsymbol{s}_{\beta})\tilde{\boldsymbol{u}}_{\beta}+\boldsymbol{b}_{\beta}.
  \end{cases} \label{ic333}
\end{align}
The first and the second equations in (48) describe the output series-paralleled converters with interactions consisting of  $\alpha$ and $\beta$ numbers of converters, respectively.
Note that \eqref{ic333} corresponds to \eqref{ic222}.
In both of these models, the circuit restrictions shown in \figref{int} have to be applied so that the models describe series or parallel connection. Thus, the same discussions as Section \ref{pair2} can be made. Since the natural numbers $\alpha$ and $\beta$ can be chosen arbitrarily, it concludes that the any output series-paralleled $n+1$ converters are described by the general model \eqref{er}.

In the previous sections, we have already shown for $n=1,\,2$ that the output series-paralleled $n$ converters are described by \eqref{er}. This statement can be proved for all $n$ by the mathematical induction based on above discussions of \eqref{ic333}. Therefore, we have proved that the output series-paralleled $n$ converters are described by \eqref{er}, for any $n\in\mathbb{N}$.

\subsection{Stabilization by PBC}

Suppose that the output series-paralleled $n$ converters
\begin{align}
  \mathcal{A}_{n}\dot{\tilde{\boldsymbol{x}}}_{n}=\{\mathcal{J}_{n}(\boldsymbol{s}_{n})
  -\mathcal{R}_{n}\}\tilde{\boldsymbol{x}}_{n}+\tilde{\boldsymbol{g}}_{n}(\boldsymbol{s}_{n})\tilde{\boldsymbol{u}}_{n}\label{demo4}
\end{align}
consist from $n$ numbers of single dc-dc converters
\begin{align}
   & \mathcal{A}_{1.m} \dot{\tilde{\boldsymbol{x}}}_{1.m}=\nonumber \\
   & \hspace{10pt}\{\mathcal{J}_{1.m} (\boldsymbol{s}_{1.m} )
  -\mathcal{R}_{1.m} \}\tilde{\boldsymbol{x}}_{1.m} +\tilde{\boldsymbol{g}}_{1.m} (\boldsymbol{s}_{1.m} )\tilde{\boldsymbol{u}}_{1.m},\label{demo5}
\end{align}
where $m\in\mathbb{N}\ \mathrm{s.t.}\ m\leq n$. Since \eqref{demo4} is written in the general form of \eqref{er}, the condition for the asymptotic stability is given from \eqref{cond} as
\begin{align}
  {\tilde{\boldsymbol{x}}_{n}}^\mathrm{T}\tilde{\boldsymbol{g}}_{n}({\boldsymbol{s}}_{n})\tilde{\boldsymbol{u}}_{n}<0,\ {\tilde{\boldsymbol{x}}_{n}}\neq0.\label{xgu4}
\end{align}
Note from the previous discussions that the series-parallel connection only alters the dissipative properties of the model. According to \eqref{param}, ${\tilde{\boldsymbol{x}}_{1.m}}$ and $\tilde{\boldsymbol{u}}_{1.m}$ are combined vertically throughout the series-parallel connection. On the other hand, $\tilde{\boldsymbol{g}}_{1.m}$ is combined diagonally. Thus, we have
\begin{align}\begin{cases}
    \tilde{\boldsymbol{x}}_{n}=\qvec{\tilde{\boldsymbol{x}}_{1.1}}{\tilde{\boldsymbol{x}}_{1.2}}{\vdots}{\tilde{\boldsymbol{x}}_{1.n}}\!,\,
    \tilde{\boldsymbol{u}}_{n}=\qvec{\tilde{\boldsymbol{u}}_{1.1}}{\tilde{\boldsymbol{u}}_{1.2}}{\vdots}{\tilde{\boldsymbol{u}}_{1.n}}\!,\,
    \\[30pt]
    \tilde{\boldsymbol{g}}_{n}=
    \begin{bmatrix}
      {\tilde{\boldsymbol{g}}_{1.1}} & 0 & \cdots & 0 \\ 0 & {\tilde{\boldsymbol{g}}_{1.2}} & ~ & \vdots \\ \vdots & ~ & \ddots & 0 \\ 0 & \cdots & 0 & {\tilde{\boldsymbol{g}}_{1.n}}
    \end{bmatrix},
  \end{cases}\label{param3}
\end{align}
and accordingly,
\begin{align}
  {\tilde{\boldsymbol{x}}_{n}}^\mathrm{T}\tilde{\boldsymbol{g}}_{n}({\boldsymbol{s}}_{n})\tilde{\boldsymbol{u}}_{n}=
  \sum^n_{m=1}{\tilde{\boldsymbol{x}}_{1.m}}^\mathrm{T}\tilde{\boldsymbol{g}}_{1.m}({\boldsymbol{s}}_{1.m})\tilde{\boldsymbol{u}}_{1.m}.\label{xgu2}
\end{align}
Controlling each converter by PBC gives
\begin{align}
  {\tilde{\boldsymbol{x}}_{1.m}}^\mathrm{T}\tilde{\boldsymbol{g}}_{1.m}({\boldsymbol{s}}_{1.m})\tilde{\boldsymbol{u}}_{1.m}<0,\ {\tilde{\boldsymbol{x}}_{1.m}}\neq0,\label{xgu3}
\end{align}
for all $m\leq n$. The condition \eqref{xgu4} is satisfied by \eqref{xgu2} and \eqref{xgu3}. Therefore, it is proved that the output series-paralleled $n$ converters are asymptotically stable by the PBC of each converter.

In conclusion of this section, we have proved that output series-paralleled dc-dc converters regulated by PBC maintains their asymptotic stability. Also, throughout the section, we have shown that any output series-paralleled converters are described by \eqref{er}. Note that the restrictions made in the discussions were only for the circuit constraints due to the output connection. It implies the robust feature of PBC that the asymptotic stability is not restricted by the diverse circuit topologies, parameters, and steady-states of the series-paralleled converters. The theoretical results are confirmed in the next section by the numerical simulation.

\section{Numerical Simulation}

In this section, the PBC of the output series-paralleled converters is examined based on numerical simulation using MATLAB/Simulink.

\begin{figure}[!t]
  \centering
  \includegraphics[scale=1.15]{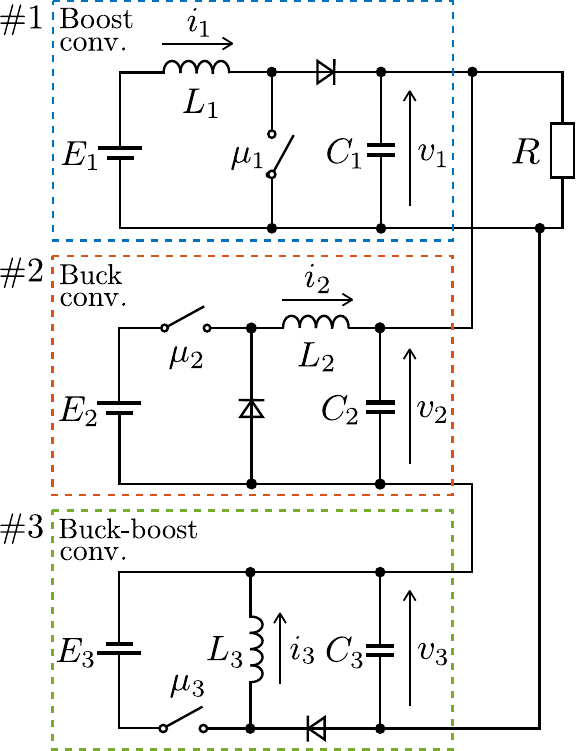}
  \caption{Boost, buck, and buck-boost converters connected in series-parallel at the output.}
  \label{bbbb}
\end{figure}

\begin{table}[ht] \caption{Parameters of dc-dc converters} \label{parameters}
  \centering
  \scalebox{1.02}{
    \begin{tabular}{cl|cl|cl}
      \hline
      \multicolumn{2}{c|}{\#1 Boost} & \multicolumn{2}{c|}{\#2 Buck} & \multicolumn{2}{c}{\#3 Buck-boost}                                             \\
      \!Parameter\!                  & \!Values\!                    & \!Parameter\!                      & \!Values\!  & \!Parameter\! & \!Values\!  \\ \hline
      $L_1$                          & 470\,$\mu$H                   & $L_2$                              & 500\,$\mu$H & $L_3$         & 330\,$\mu$H \\
      $C_1$                          & 10\,$\mu$F                    & $C_2$                              & 33\,$\mu$F  & $C_3$         & 20\,$\mu$F  \\
      $E_1$                          & 18\,V                         & $E_2$                              & 40\,V       & $E_3$         & 24\,V       \\
      $k_1$                          & 0.02                          & $k_2$                              & 0.3         & $k_3$         & 0.02        \\
      \hline\end{tabular}}
  \vspace{15pt}

  \caption{Initial state of dc-dc converters} \label{initial}
  \centering
  \scalebox{1.02}{
    \begin{tabular}{cl|cl|cl}
      \hline
      \multicolumn{2}{c|}{\#1 Boost} & \multicolumn{2}{c|}{\#2 Buck} & \multicolumn{2}{c}{\#3 Buck-boost}                                        \\
      \!Variable\!                   & \!State\!                     & \!Variable\!                       & \!State\! & \!Variable\! & \!State\! \\ \hline
      $i_1$                          & 1.4\,A                        & $i_2$                              & 1.3\,A    & $i_3$        & 2.8\,A    \\
      $v_1$                          & 10\,V                         & $v_2$                              & 16\,V     & $v_3$        & 12\,V     \\
      \hline\end{tabular}}
  \vspace{15pt}

  \caption{Desired state of dc-dc converters} \label{desired}
  \centering
  \scalebox{1.02}{
    \begin{tabular}{cl|cl|cl}
      \hline
      \multicolumn{2}{c|}{\#1 Boost} & \multicolumn{2}{c|}{\#2 Buck} & \multicolumn{2}{c}{\#3 Buck-boost}                                               \\
      \!Variable\!                   & \!State\!                     & \!Variable\!                       & \!State\! & \!Variable\!        & \!State\! \\ \hline
      $i_{1\mathrm{d}}$              & 1.950\,A                      & $i_{2\mathrm{d}}$                  & 2.025\,A  & $i_{3\mathrm{d}}$   & 3.375\,A  \\
      $v_{1\mathrm{d}}$              & 36\,V                         & $v_{2\mathrm{d}}$                  & 20\,V     & $v_{3\mathrm{d}}$   & 16\,V     \\
      $\mu_{1\mathrm{d}}$            & 0.5                           & $\mu_{2\mathrm{d}}$                & 0.5       & $\mu_{3\mathrm{d}}$ & 0.4       \\
      \hline\end{tabular}}
\end{table}

\begin{figure}[!t]
  \centering
  \includegraphics[width=1\columnwidth]{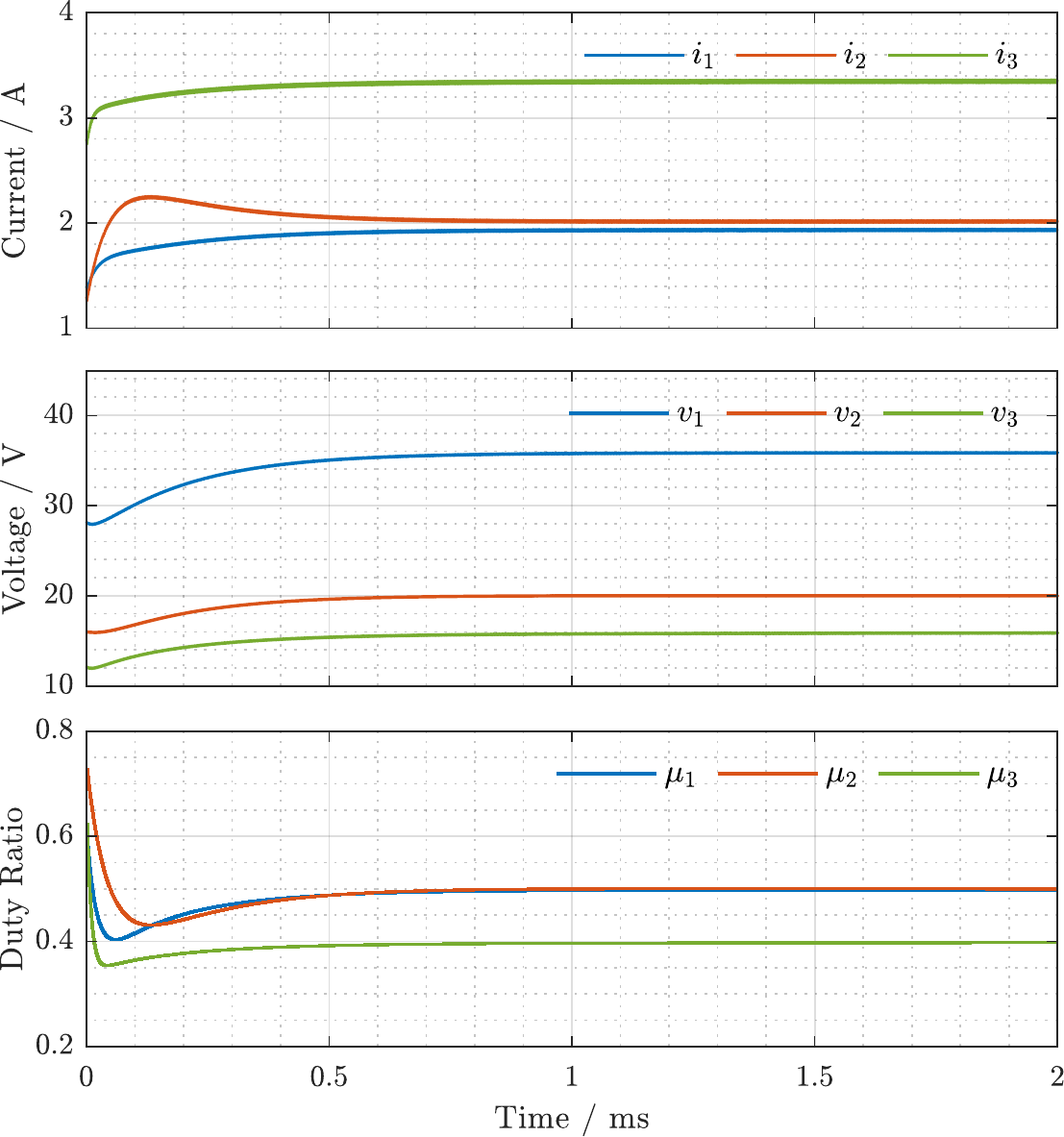}
  \caption{Numerical waveforms of output series-paralleled converters for the ideal case with no disturbances.}
  \label{waveforms}
\end{figure}

\subsection{Setups for Simulation}

The simulated circuit is shown in \figref{bbbb}. It is composed of boost, buck, and buck-boost converters. The subscripts `1', `2', and `3', correspond to the boost, buck, and buck-boost converter, respectively. The circuit parameters for the simulated circuit are shown in \tabref{parameters}. Here, the parameters are chosen to be diverse. The parameter $k$ is the feedback control gain set for the PBC of each converter. The
load resistance is $R = 12\,\Omega$. The switches and the diodes are considered ideal.

As for control, the PBC explained in section II are applied to each converter. The duty ratios for the switches of boost, buck, and buck-boost converters are governed by \eqref{boostpbc}, \eqref{buckpbc}, and \eqref{buckboostpbc}, respectively. The calculated duty ratios are converted to the switching signal by PWM at 1\,MHz. The sampling frequency of the state variables is also set at 1\,MHz, and the calculation of the duty ratios is executed instantly. The feedback control scheme follows the schematic diagram shown in \figref{pbcblockdiagram}.

The initial state and the desired state of the simulated circuit are given in \tabref{initial} and \tabref{desired}, respectively. The desired state is determined for the converters to have various voltage, current, and power levels at the steady-state. In the simulation, we first consider the ideal case that there is no disturbance in the system, to confirm the asymptotic stability numerically. Then, we also consider the cases that there are perturbations or fluctuations at the input voltage source or the output load, in order to examine the robustness of the proposed control method. These simulation scenarios are based on the previous works in \cite{Sira-Ramirez1997, Ortega1998}.

\subsection{Results}
\subsubsection{Ideal Case}
The simulated waveforms of the output series-paralleled converters in the ideal case are shown in \figref{waveforms}. It can be seen from \figref{waveforms} that the waveforms converge to the desired state shown in \tabref{desired}. Therefore, PBC was numerically confirmed to achieve the asymptotic stability of the series-paralleled boost, buck, and buck-boost converters. It was also confirmed that the diversity in each converter did not violate the asymptotic stability. This implies the robust feature of PBC that it allows the diverse voltage, current, and power levels among various circuit parameters in the series-parallel connection.

\begin{figure}[!t]
  \centering
  \includegraphics[width=1\columnwidth]{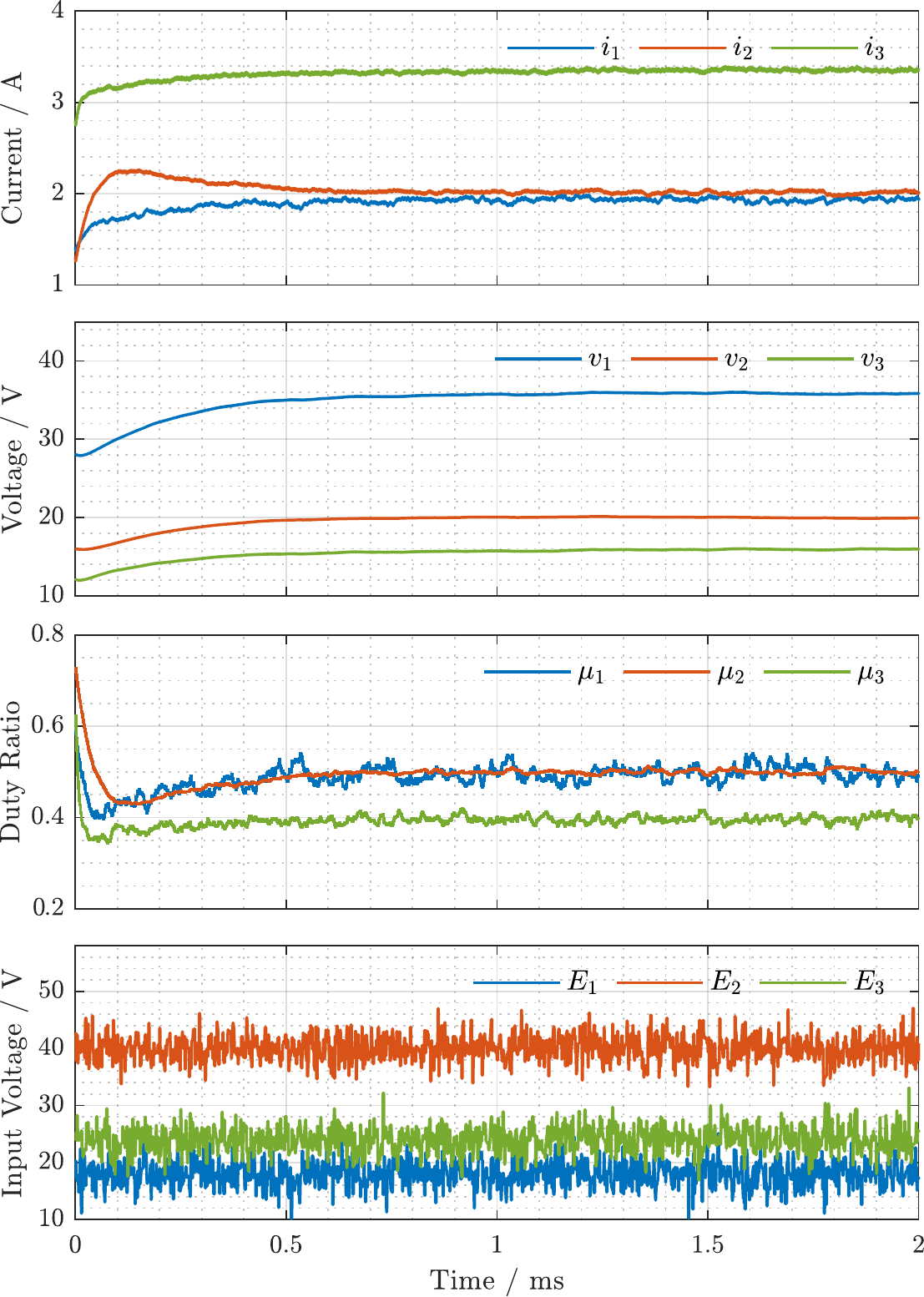}
  \caption{Numerical waveforms of output series-paralleled converters with random perturbation input.}
  \label{waveforms1}
\end{figure}

\subsubsection{Perturbation at Input Voltage Source}

For this simulation, a random perturbation was added to the input voltage sources $E_1$, $E_2$, and $E_3$. The maximum peak-to-peak magnitude of the perturbation was chosen to be approximately 10\,V. This is 55.6\,\%, 41.7\,\%, and 25.0\,\% of the values of $E_1$, $E_2$, and $E_3$, respectively.

\figref{waveforms1} shows the simulated waveforms of the output series-paralleled converters with the perturbation inputs. The bottom graph shows the perturbed input voltage. As can be seen, the proposed PBC achieves the stability of the system around the desired state of \tabref{desired}. The instantaneous errors of the input currents and the output voltages from their desired states are less than 4.1\,\% and 1.9\,\%, respectively. The variations of computed duty ratios are within 5.0\,\% of the total range of the closed interval [0,1]. Hence, the error in the system is found to be significantly reduced compared to the input perturbation. PBC exhibited high robustness with regards to the perturbed inputs. It is worth noting that the low-pass nature of the converter itself also contributes to the reduction of the error as well as the control.

\begin{figure}[!t]
  \centering
  \includegraphics[width=1\columnwidth]{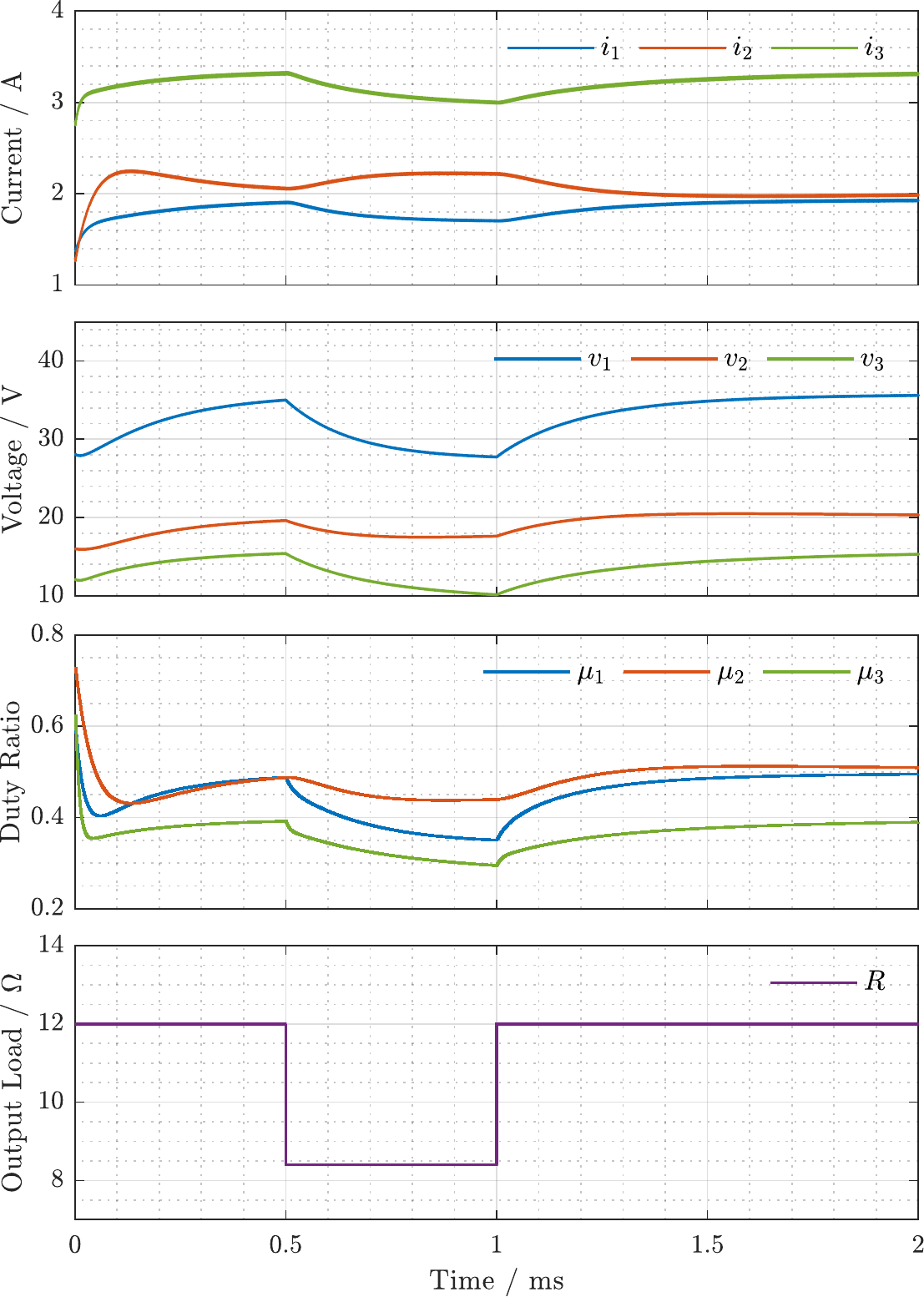}
  \caption{Numerical waveforms of output series-paralleled converters with temporary load fluctuation.}
  \label{waveforms2}
\end{figure}

\subsubsection{Temporary Fluctuation of Output Load}

Unknown load resistance variations generally affect the behavior of the closed-loop performance of the controlled converter \cite{Ortega1998}. In this simulation, temporary fluctuation of the output load resistance $R$ was given. The obtained waveforms are shown in \figref{waveforms2}. As seen in the bottom graph, the output load resistance temporarily decreases to 70\,\% of its nominal value. It is confirmed that the PBC restores its asymptotic stability after the change, converging towards the desired state of \tabref{desired}. The temporary change of the load did not violate the stability nor the proper steady-state operation of the output series-paralleled converters regulated by PBC.

\subsection{Discussions}

In this section, we numerically confirmed the stability and the robustness of the output series-paralleled converters regulated by PBC. The simulated results showed correspondence to the theoretical discussions in that PBC achieved the stabilization of output series-paralleled converters with various circuit topologies, parameters, and steady-states. The robustness of the proposed control was also examined through the perturbations. Although the obtained waveforms highly depend on the settings of the circuit parameters, desired states, and disturbances, the overall features of stability and robustness are likely to remain widely, since our consideration does not require any specific settings of the system. These features correspond to the experimental result in \cite{Murakawa2020}, where PBC was shown to asymptotically stabilize the paralleled boost and buck converters with various parameters and steady-states.

\section{Conclusion}

In this paper, we examined the asymptotic stability of the output series-paralleled dc-dc converters regulated by PBC. We introduced the dc-dc converter model with the additional variable representing the output interaction to describe the series-parallel connection of the converters. Then, the output series-paralleled converters were shown to be reclassified into the general dc-dc converter model. This feature enabled us to discuss the asymptotic stability of the series-paralleled converters based on the condition obtained from the general converter model. Consequently, the converters regulated by PBC were proved to maintain their stability at the output series-parallel connection. The theoretical result was verified in the numerical simulation of series-paralleled boost, buck, and buck-boost converters. The stable and robust features of the proposed control were confirmed in the simulation. PBC allowed diverse circuit topologies, parameters, and steady-states of the output series-paralleled dc-dc converters. Our contribution is able to theoretically support the previous numerical and experimental studies for the series-paralleled converters regulated by PBC, justifying the further extension of the system.

\ifCLASSOPTIONcaptionsoff
  \newpage
\fi

\end{document}